\documentclass[useAMS,usenatbib,usegraphicx]{mn2e}

\usepackage{aas_macros}

\usepackage{multirow}
\usepackage[caption=false]{subfig}
\usepackage{pdfpages}

\title[Modelling the chemistry of star forming filaments]{Modelling the chemistry of star forming filaments -- I. H$_2$ and CO chemistry}
  \author[D. Seifried et al.]
  {D.~Seifried,$^{1}$\thanks{seifried@ph1.uni-koeln.de}, S. Walch,$^{1}$ \\
  $^1$I. Physikalisches Institut, Universit\"at zu K\"oln, Z\"ulpicher Str. 77, 50937 K\"oln, Germany}

\date{Released 2015}

\pagerange{\pageref{firstpage}--\pageref{lastpage}} \pubyear{2015}

\begin{document}

\label{firstpage}

\maketitle

\begin{abstract}

We present simulations of star forming filaments incorporating on of the largest chemical network used to date on-the-fly in a 3D-MHD simulation. The network contains 37 chemical species and about 300 selected reaction rates. For this we use the newly developed package KROME \citep{Grassi14}. We combine the KROME package with an algorithm which allows us to calculate the column density and attenuation of the interstellar radiation field necessary to properly model heating and ionisation rates. Our results demonstrate the feasibility of using such a complex chemical network in 3D-MHD simulations on modern supercomputers. We perform simulations with different strengths of the interstellar radiation field and the cosmic ray ionisation rate. We find that towards the centre of the filaments there is gradual conversion of hydrogen from H to H$_2$ as well as of C$^+$ over C to CO. Moreover, we find a decrease of the dust temperature towards the centre of the filaments in agreement with recent HERSCHEL observations.
\end{abstract}


\section{Introduction}

Modelling the chemical evolution of the gas during the process of star formation on various scales is a numerically and theoretically challenging task. On the one hand, it requires to solve a large set of reaction equations with sufficient accuracy, which is computationally extremely demanding. On the other hand in depth knowledge about the necessary reactions and their corresponding rates is required, which involves laboratory work and quantum mechanical calculations.

Nonetheless, the chemical state of the gas and the associated cooling and heating processes are required to properly model the thermodynamical evolution of the gas and to make accurate predictions via synthetic observations. In particular in the latter case it is common to use fixed conversion factors to e.g. obtain the number density of a particular molecular species from the gas density which, however, is a severe oversimplification and might not be accurate enough.

Several authors have incorporated reduced chemical networks in their 3D, magneto-hydrodynamical (MHD) simulations \citep[e.g.][]{Clark13,Smith14,Walch15,Hocuk16}. The newly designed chemistry package KROME \citep{Grassi14} is a versatile package which allows the user to incorporate chemistry in MHD simulations in a very efficient manner and simultaneously guarantees the freedom to choose \textit{any} desired network as well as its associated cooling and heating processes. It is therefore a good choice to model the complex chemical evolution of star forming regions.

Recently, the importance of filamentary structures in star forming clouds has been re-emphasized \citep[e.g.][]{Andre10}. These filaments result from the interaction of gravity and turbulence and harbour pre- and protostellar cores \citep[for recent numerical works see e.g.][]{Gomez14,Smith14,Kirk15}. Recent observations of molecular line and dust emission provide a wealth of information about the properties of these filaments, like e.g. the thermodynamical state, the typical extension, and fragmentation of the filaments. Numerical simulations, however, often lack the detailed description of the chemical state and can thus be compared only indirectly to the observations. This and the fact that filaments appear to be ubiquitous in the star formation process make them an optimal target to test the feasibility of using a detailed chemical network in fully self-consistent 3D-MHD simulations.

The outline of the paper is as follows: In Section~\ref{sec:IC} we briefly discuss the initial conditions, in Section~\ref{sec:chem} we describe the employed chemical network and heating and cooling processes. In Section~\ref{sec:results} we describe the time evolution of a fiducial run and study the impact of the radiation field on the chemical and thermal evolution of the filaments. In Section~\ref{sec:dis} we compare with observations and give details about the computational cost before we  summarize our results.

\section{Initial conditions and FLASH solver}
\label{sec:IC}

The simulations presented here use the same initial conditions and numerical methods -- except the chemistry and cooling/heating processes -- as those presented in \citet{Seifried15}. Here we briefly summarize the main points. The simulations are performed with the astrophysical code FLASH 4.2.2 \citep{Dubey08}. We solve the equations of ideal MHD using a maximum resolution of 40.3 AU. The Poisson equation for gravity is solved using a multipole method based on a Barnes-Hut tree \citep{Wunsch15}.

The simulated filaments have an initial width of about 0.1 pc following a Plummer-like density profile along the radial direction. The filaments have a length of 1.6 pc and a mass per unit length of 75 M$_{\sun}$/pc. This corresponds to about three times the critical mass per unit length \citep{Ostriker64} of
\begin{equation}
   (M/L)_\rmn{crit} = \frac{2 c_\rmn{s}^2}{G} \, ,
\end{equation}
which is why the filaments are unstable against collapse along the radial direction and subject to subsequent fragmentation. The initial central density of the filaments is $3\times10^{-19}$ g cm$^{-3}$.

Recent observations have shown that filaments have magnetic fields which can be directed either along or perpendicular to the filament \citep[e.g.][]{Sugitani11,Li13,Palmeirim13}, which is why we test both magnetic field orientations. We take the initial magnetic field strength in the centre of the filament to be 40 $\mu$G, in agreement with recent observations \citep[e.g.][]{Sugitani11}. The field is uniform in strength for the perpendicular case and declines outwards in the parallel case proportional to $\rho^{0.5}$. In addition, we superimpose a transonic turbulent velocity field \citep[for more details see][]{Seifried15}.

The strength of the interstellar radiation field (ISRF) and the cosmic ray ionisation rate (CRIR) influence the reaction rates as well as the heating of dust, which is taken into account by the chemistry package (see Online Material). Since, however, the strength of the ISRF and the CRIR can vary locally in our Galaxy, we perform simulations with different values. In our fiducial run we apply a CRIR of $1.3 \times 10^{-17}$ s$^{-1}$ \citep[e.g.][]{Vastel06} and an ISRF corresponding to 1.7 times the Habing flux \citep[G$_0$ = 1.7,][]{Draine78}. Since for the CRIR also significantly higher values are found \citep[e.g.][]{Caselli98,Ceccarelli11}, we perform a second pair (perpendicular and parallel magnetic fields) of simulations with a CRIR of $1 \times 10^{-16}$ s$^{-1}$ and G$_0$ = 1.7. In the third set of simulations we additionally increase the ISRF by a factor of 5, i.e. G$_0$ = 8.5. Hence, in total we have a set of six simulations with different (initial) conditions.

\section{The chemical network}
\label{sec:chem}

In order to model the chemical evolution of the gas, we use the KROME package \citep{Grassi14}, which in principle allows to model any desired chemical network. The network used contains 37 species and 287 reactions including different forms of hydrogen and carbon bearing species like H$^+$, H, H$_2$, C$^+$, C, and CO but also more complex species like e.g. HCO$^+$, H$_2$O, or the cosmic ray tracer H$_3^+$. In particular, the network allows for a very detailed description of the formation of CO and H$_2$ (also including the formation of H$_2$ on dust). We assume that all elements heavier than He are depleted with respect to their cosmic abundances using values given by \citet{Flower05} typical for dense molecular gas. Initially all elements are in their atomic form. More details about the used species, reactions, and initial abundances are provided in the Online Material. In addition, we benchmark our implementation (network and radiative transfer) against the KOSMA-$\tau$ PDR code for two tests presented in \citet{Roellig07} and find very good agreement.

The strength of the ISRF and the CRIR are needed to model the chemical evolution as well as the heating and cooling processes properly. In particular, the ISRF experiences an attenuation when travelling through the gas. We calculate the attenuation in the FLASH code with the TreeCol algorithm \citep{Clark12,Wunsch15}. For each cell in the computational domain we compute the total (H + H$_2$), H$_2$, and the CO column densities in 48 directions. We take the geometrical average and obtain the mean column densities, N$_\rmn{H,tot}$, N$_\rmn{H_2}$, and N$_\rmn{CO}$, as well as the visual extinction, $A_V$, and the self-shielding factors for H$_2$ and CO photodissociation \citep[][see also section 2.2.1 and 2.2.2 in \citet{Walch15} for a more detailed technical description]{Glover10}. The attenuation of the ISRF also affects the photoelectric (PE) heating due to dust particles, which is taken into account by scaling the strength of the ISRF with G$_0$ $\times$ exp(-2.5 $A_V$).

The cooling rate due to line emission of CO is provided by K. Omukai in tabulated form based on the results of \citet{Neufeld93} and \citet{Omukai10}. The actual rate in each cell depends on N$_\rmn{CO}$ and the local gas density and gas temperature, $T_\rmn{gas}$ . We also use these tabulated rates to calculate the cooling due to $^{13}$CO and C$^{18}$O by downscaling N$_\rmn{CO}$ and the according cooling rates by 69 and 557, respectively \citep{Wilson99}.

Finally, for calculating the dust temperature, $T_\rmn{dust}$, we take into account the transfer of energy from gas to grains via collisions, which also affects $T_\rmn{gas}$ itself, heating of the grains due to the ISRF, and cooling of dust grains via black-body radiation. In each timestep, we calculate the equilibrium dust temperature for all three processes after the chemistry update is done. We emphasize that a more detailed description of all cooling and heating processes used and their coupling to the ISRF and the CRIR is provided in the Online Material.

\section{Results}
\label{sec:results}

\subsection{Time evolution of a fiducial run}

Before we start the different simulations, we evolve the chemistry in each cell for 500 kyr, during which the hydrodynamical evolution is \textit{turned off}, i.e. only $T_\rmn{gas}$, $T_\rmn{dust}$, and the chemical composition are updated. Since at densities of about 10$^5$ cm$^{-3}$ the typical timescale for H$_2$ formation on dust grains is of the order of 100 kyr, 500 kyr are sufficient to reach a rough chemical equilibrium. After this initial relaxation phase, each simulation is followed for \mbox{300 kyr}.

\begin{figure}
 \includegraphics[width=\linewidth]{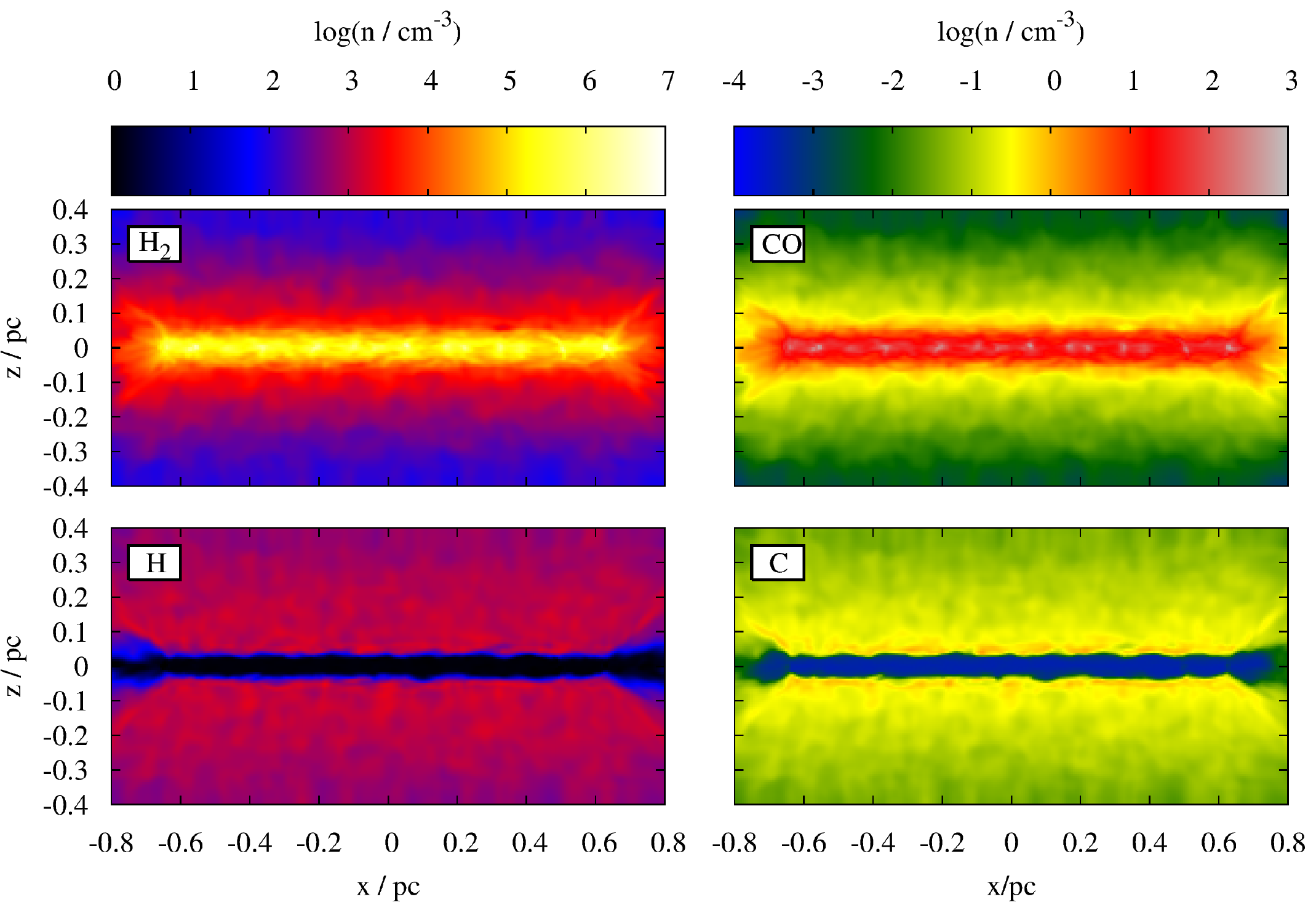}
 \caption{Spatial distribution of  H, H$_2$, C, and CO at the end of the simulation with a parallel magnetic field, G$_0$ = 1.7, and a CRIR of 1.3 $\times$ 10$^{-17}$ s$^{-1}$ along a slice through the centre of the filament.}
 \label{fig:species}
\end{figure}

In the following, we present the time evolution of the run with a parallel magnetic field, G$_0$ = 1.7, and a CRIR of 1.3 $\times$ 10$^{-17}$ s$^{-1}$. In Fig.~\ref{fig:species} we plot the spatial distribution of H, H$_2$, C, and CO at the end of the simulation, which reveals some differences in the radial distribution. Whereas H$_2$ and CO are concentrated towards the centre of the filament, H and C are more extended. Moreover, there is a clear impact of the initial turbulence field on the distribution of the chemical species, causing strong local variations.

\begin{figure}
 \includegraphics[width=\linewidth]{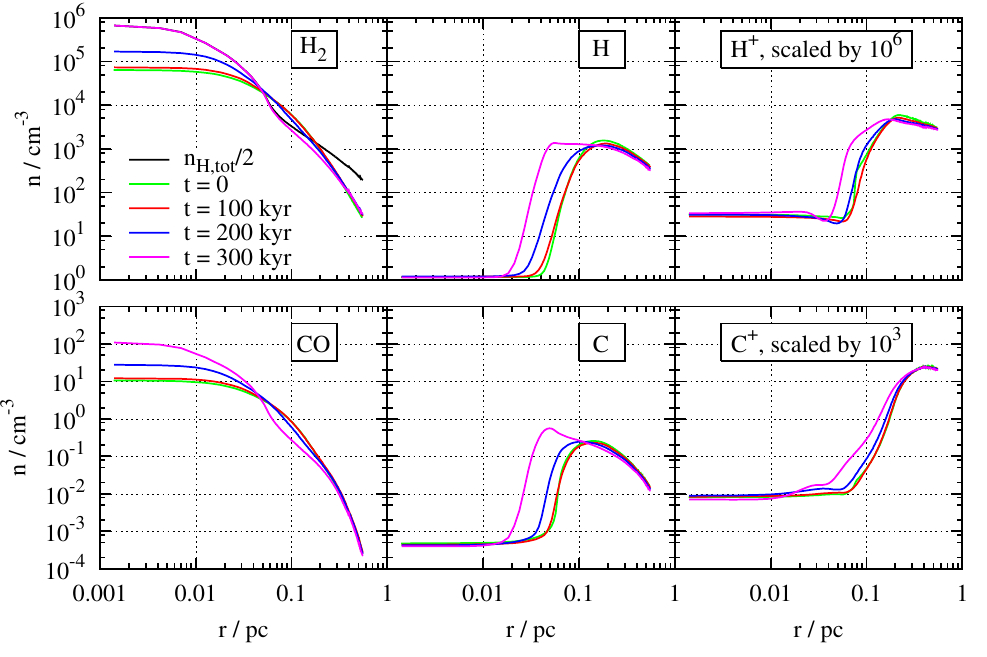}
 \caption{From left to right: Radial dependence of the density of  H$_2$, H, H$^+$ (top row), and CO, C, and C$^+$ (bottom row) at \mbox{t = 0, 100, 200, and 300 kyr} after the start of the same simulation as in Fig.~\ref{fig:species}. Note that at t = 0, the chemical distribution corresponds to that after an initial relaxation phase of 500 kyr. In order to use the same $y$-axis, $n_\rmn{H^+}$ and $n_\rmn{C^+}$ are scaled up by a factor of 10$^6$ and 10$^3$, respectively. The black line in the upper left panel shows the total hydrogen density divided by two.}
 \label{fig:time}
\end{figure}

The radial density profiles of H$_2$, H, H$^+$, CO, C, and C$^+$ at 4 different times are shown in Fig.~\ref{fig:time}. In order to smooth out local fluctuations, we have averaged the densities along the major axis of the filament.
The constant increase of the central number densities of H$_2$ and CO over time is caused by the contraction of the filament along the radial direction. On the other hand, the abundance of H, H$^+$, C, and C$^+$ remain almost constant over time as they recombine when the filament becomes denser. In the upper left panel we also show the number density of all H atoms (divided by two in order to be comparable to $n_\rmn{H_2}$), which represents the shape of the mass density profile. In the centre of the filament most of the hydrogen is bound in H$_2$, only outside $\sim$ 0.1 pc, where the black and purple line start to differ, hydrogen mainly occurs in atomic form. Hence, there is a gradual conversion of H to H$_2$ towards the centre of the filament as well as of C$^+$ over C to CO, with $n_\rmn{C}$ exceeding $n_\rmn{CO}$ at $\sim$ 0.1 pc as well. This agrees with the picture obtained from Fig.~\ref{fig:species} that H and C envelop H$_2$ and CO and is in good agreement with theoretical predictions of PDRs \citep[e.g.][]{Dishoeck88}. We note that a similar result was also found by \citet{Clark13}.

\subsection{Impact of the ISRF and the CRIR}

\begin{figure*}
 \includegraphics[width=0.9\linewidth]{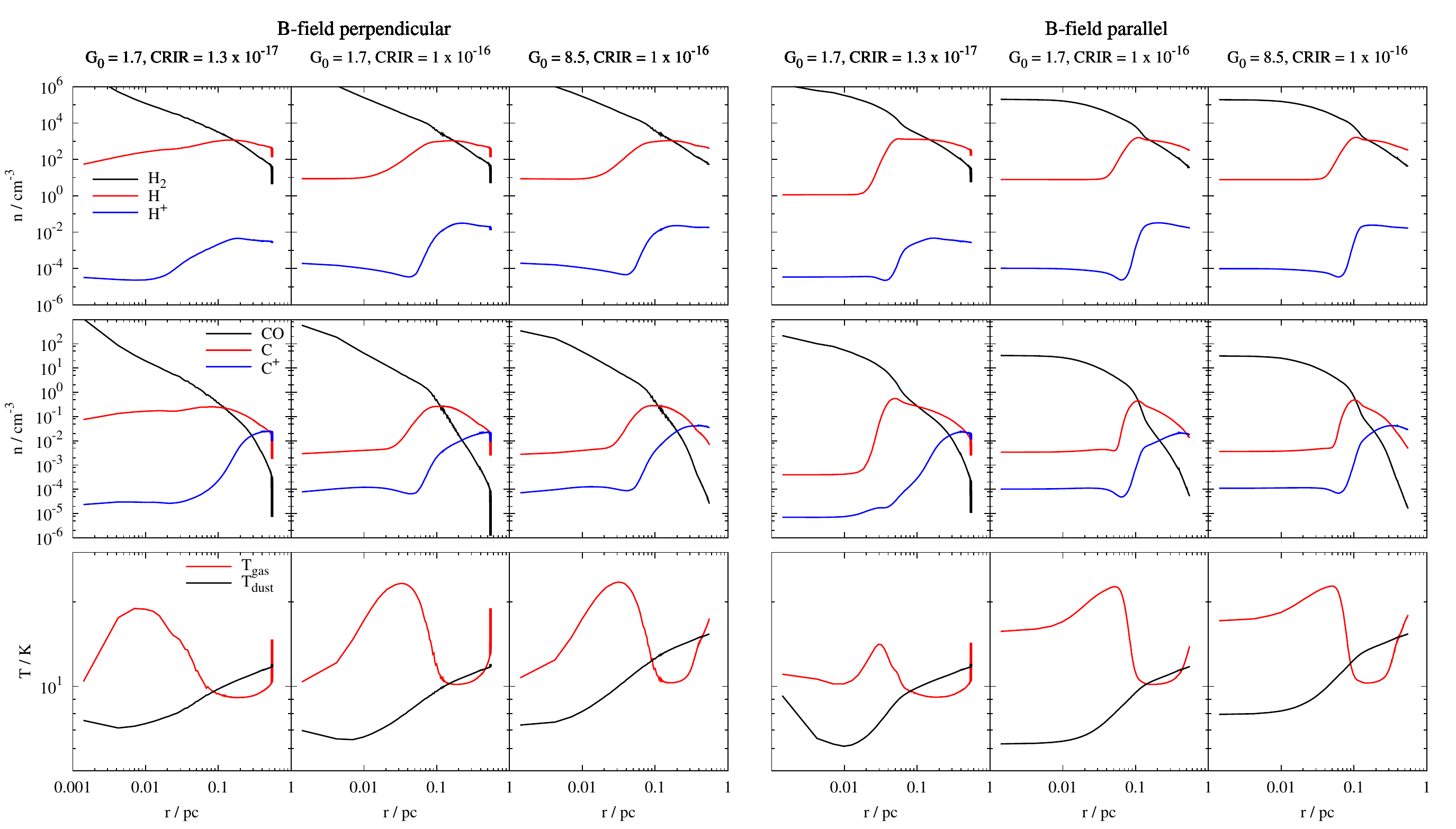}
 \caption{Radial dependence of H$_2$, H, H$^+$ (top row), CO, C, C$^+$ (middle row) as well as the gas and dust temperature (bottom row) at the end of each simulation. Left panel: runs with a perpendicular magnetic field. Right panel: runs with a parallel magnetic field.}
 \label{fig:radial}
\end{figure*}

Next, we study the influence of the ISRF and CRIR on the properties of the filaments. In Fig.~\ref{fig:radial} we plot the radial dependence of H$_2$, H, and H$^+$ (top row), as well as of CO, C, and C$^+$ (middle row) and of $T_\rmn{gas}$ and $T_\rmn{dust}$, (bottom row) at the end of each of the six simulations. On the left we show the runs with a perpendicular magnetic field, on the right runs with a parallel magnetic field. As already seen in Fig.~\ref{fig:time}, we find a gradual conversion of H to H$_2$, and C$^+$ over C to CO towards centre of the filaments for all runs.

Increasing the CRIR results in a somewhat higher ionisation reflected by the abundances of H$^+$ and C$^+$. Moreover, at high densities $n_H$ is around a few times 1 cm$^{-3}$ with somewhat higher values for higher CRIR and thus in good agreement with theoretical predictions \citep[e.g.][]{Tielens05}. Also $T_\rmn{gas}$ shows a slight increase with increasing CRIR, which is due to the larger amount of energy released by various dissociation reactions caused by cosmic rays.

Increasing the strength of the ISRF only marginally affects the chemical composition of the gas. However, $T_\rmn{gas}$ and $T_\rmn{dust}$ are increased by a few Kelvin, which is most likely due to the increased PE heating. Interestingly, we find that for all runs $T_\rmn{dust}$ decreases towards the centre of the filaments. We attribute this to the progressive attenuation of the ISRF -- mainly responsible for dust heating -- towards the centre of the filament \citep[see also][]{Clark13}. The gas temperature, however, shows a significant increase between 0.01 pc and 0.1 pc. Investigating the dynamics of the filaments, we find that this corresponds to an accretion shock where the infalling gas is decelerated.

As shown in \citet{Seifried15} the magnetic field has a strong impact on the dynamical evolution of the filament, which is partly also reflected in the chemical composition. In particular the two runs with G$_0$ = 1.7 and a CRIR of 1.3 $\times$ 10$^{-17}$ s$^{-1}$ show a somewhat different chemical composition with respect to atomic hydrogen and carbon. A more significant difference can be seen in the gas temperature, which as mentioned before, is tightly linked to the infall velocities, which in turn depends on the configuration of the magnetic field \citep{Seifried15}.

\section{Discussion and conclusions}
\label{sec:dis}

\subsection{Physical interpretation}

As shown in Fig.~\ref{fig:radial}, $T_\rmn{dust}$ decreases towards the centre of the filament in all runs, which was recently also found in observations \citep{Arzoumanian11,Palmeirim13,Li14}. From the averaged quantities (Figs.~\ref{fig:time} and~\ref{fig:radial}) we can now obtain the polytropic index $\gamma$ of the dust by assuming a polytropic relation $T_\rmn{dust} \propto \rho^{\gamma - 1}$.  We note that we rather use a relation between the $T_\rmn{dust}$ (instead of $T_\rmn{gas}$) and $\rho$ in order to be able to directly compare our results to HERSCHEL observations: by visual inspection of the $T_\rmn{dust}$ vs. $\rho$ plots (see Online Material), we find $\gamma$ to be in a range from 0.9 to 0.95, which is in good agreement with \citet{Palmeirim13}. We again emphasize that $T_\rmn{gas}$ and $T_\rmn{dust}$ are markedly different, which requires independent measurements for both quantities in observations. We attribute this to the fact that the collisional coupling between gas and dust at $n \sim 10^5 - 10^6$ cm$^{-3}$ is not yet strong enough to assure similar temperatures, although our results suggest that $T_\rmn{dust}$ and $T_\rmn{gas}$ approach each other towards the center of the filaments.

By inspecting Fig.~\ref{fig:radial} it can also be seen that the ratio $R = \frac{n_{\rmn{CO}}}{n_\rmn{H_2}}$ is not constant along the radial direction. There is a strong decline of $R$ with increasing radius of about 2 orders of magnitude, which is due to the fact that the formation of CO happens at somewhat smaller radii (i.e. higher gas column densities) than that of H$_2$. This variation in $R$ most likely also affects the value of the X-factor, which is used to convert CO luminosities into H$_2$ column densities. Our work therefore suggests that caution is recommended when using CO line intensities to obtain total gas masses. Furthermore, varying CRIR and ISRF mainly affects $R$ at radii larger than 0.05 pc. In the centre of the filament, however, $R$ is comparable for all our runs with a value around $1.5 \times 10^{-4}$, and thus in good agreement with observations \citep[see][for a recent review]{Tielens13} despite our simplified depletion description. We note that the variation of $R$ over time is rather small, independent of the considered run.

Finally, we emphasise that also abundances of other species like H$_3^+$, H$_2$O, OH, or CH included in the network seem to be in reasonable agreement with observations \citep[e.g.][]{Tielens13}. This is a significant step forward compared to smaller chemical networks used in previous works \citep[e.g.][]{Clark13,Smith14,Walch15}. We postpone a more in-depth discussion of these species to a subsequent paper.

\subsection{Numerical cost/Performance}

The simulations were carried out on computing nodes with 2 - 5 GB memory per CPU, which corresponds to state-of-the-art computing nodes at modern supercomputing facilities (see Acknowledgements for which facilities were used). In order to estimate the increase of computational power required, we compare our runs with an isothermal ($T$ = 15 K) simulation without a chemical network, which is also evolved for 300 kyr. The slightly different thermodynamical evolution of the filaments in runs with and without the network results in slightly different timesteps as well as different numbers of cells in the simulation domain, which cannot be avoided. However, these differences affect the cost estimates only marginally. Overall, we find that the inclusion of the network (including heating and cooling processes) increases the computation costs per timestep by roughly a factor of 7. However, tabulating the reaction rate coefficients, which avoids the repeated evaluation of the exponential function, reduces the cost increase to about a factor of 5.

Altogether we argue that -- considering the ever increasing computing power, the higher accuracy as well as the significant gain of additional physical and chemical information -- it seems feasible and necessary for the future to perform simulations with large chemical networks.

\subsection{Conclusions and outlook}

We present first results of simulations of star forming filaments using one of the largest chemical network ever applied in 3D-MHD simulations. For this we combine the versatile chemistry package KROME \citep{Grassi14} with the TreeCol algorithm \citep{Clark12,Wunsch15}, which allows us to self-consistently calculate the optical depth and shielding parameters in the simulation domain, which in turn is required to properly determine dissociation and photoionization reactions. The network contains 37 species and 28.pdf7 thoroughly selected reactions. We show that in terms of memory consumption such simulations are feasible on modern supercomputers. Despite an increase of computation costs by a factor of $\sim$ 5 -- 7, we argue that the significant gain of physical and chemical information justifies the usage of large chemical networks in future simulations.

Concerning the chemical composition as well as the thermodynamical behaviour of the simulated filaments, the results appear to be promising and in good agreement with observational results. In particular, we show that there is a gradual conversion of H to H$_2$ as well as of C$^+$ over C to CO towards the centre of the filaments. Moreover, we find that the dust temperature $T_\rmn{dust}$ decreases towards the centre of the filament following a polytropic relation $T_\rmn{dust} \propto \rho^{\gamma - 1}$ with $\gamma \simeq 0.9 - 0.95$ in agreement with recent observations \citep{Arzoumanian11,Palmeirim13,Li14}. Furthermore, we show that the dust and gas temperature can be markedly different even at densities of $n \sim 10^5 - 10^6$ cm$^{-3}$.

This work paves the way for many future applications: The detailed chemical modelling will allow us to produce synthetic observations of several molecular lines as well as of continuum emission. Filament widths obtained from these synthetic observations allow us to study how well a projected (2D--) width matches the actual (3D--) width of the filament obtained directly from the simulation data. Furthermore, the simulations can improve our understanding of the X-factor required for the mass determination of gaseous objects. Moreover, the network will allow us to study in detail the evolution of other molecules like H$_2$O or the cosmic ray tracer H$_3^+$, which, due to space limitations, we could not discuss here. We also note that we plan to include even more complex chemical networks. In particular, the inclusion of nitrogen, a constituent of a number of important observational tracers, is of large interest. Finally, we plan to improve our treatment of CO freeze-out in order to get a better handle on the X-factor.

\section*{Acknowledgements}

The authors like to thank the referee for the comments which helped to significantly improve the paper. Furthermore, the authors thank T. Grassi and S. Bovino for the development of the KROME chemistry package and the useful discussions, as well as S. Glover for discussions on the implementation of a dust cooling routine. The authors acknowledge funding by the Bonn-Cologne Graduate School as well as the Deutsche Forschungsgemeinschaft (DFG) via the Sonderforschungsbereich SFB 956 \textit{Conditions and Impact of Star Formation} and the Schwerpunktprogramm SPP 1573 \textit{Physics of the ISM}. The simulations were performed on JURECA at the Computing Center J\"ulich. The FLASH code was developed partly by the DOE-supported Alliances Center for Astrophysical Thermonuclear Flashes (ASC) at the University of Chicago.

\label{lastpage}



\section*{Supplementary material}

# Online Material: Modelling the chemistry of star forming filaments – I. $H_2$ and CO chemistry

2 March 2016

## 1 CHEMICAL SPECIES, ABUNDANCES, AND REACTION RATES

For the simulations presented in this work we use the precompiled chemical network named `react_COthin` contained in the KROME package (Grassi et al. 2014, but see also www.kromepackage.org). It contains in total 37 species, which we list in Table 1, as well as a thorough selection of 287 chemical reactions in order to accurately describe the detailed evolution of various molecules like $H_2$ and CO.

Here we briefly recapitulate the main reactions contained. The network contains a reaction to model the formation of $H_2$ on dust grains, which depends on the dust temperature, which in turn we obtain self-consistently from the simulations (see Section 2.2). Furthermore, the network includes a selection of photochemical reaction rates describing the photodissociation and ionization of various molecules and atoms due to the incident interstellar radiation field (ISRF, see Table B2 in the appendix of Glover et al. 2010, for a compilation of all photochemical reaction rates included in the network). In our simulations the reaction rates listed in Glover et al. (2010) are scaled by $G_0/1.69$, where $G_0$ is the strength of the ISRF in terms of the Habing flux (Draine 1978, but see also Section 2 of the paper). The network also contains reactions due to cosmic rays taken from the KIDA database (http://kida.obs.u-bordeaux1.fr/), which depend on the value of the cosmic ray ionisation rate (CRIR). Additional reaction rates are compiled in from several sources like the KIDA database, the UMIST database (http://udfa.ajmarkwick.net/), as well as a number of authors (O'Connor et al. 2014; Kingdon & Ferland 1996; Glover & Jappsen 2007; Glover et al. 2010). Since the network is by far too extensive to explain all its details, in Fig. 1 we give a graphical representation of a selection of reaction channels. For reasons of readability not all reactions[1] are included in the plot. However, on request we will provide the entire network to the interested reader.

As mentioned in Section 3 of the paper, we take into account the depletion/freeze-out of carbon, oxygen, and silicon onto dust grains in the cold and dense interstellar medium in a simplified manner. In order to do so, we reduce the overall abundances of the different elements with respect to their cosmic abundance following Flower et al. (2005, see their Table 1). In Table 2 we list the mass fractions as well as the fractional abundances $\frac{n_x}{n_{H,tot}}$, where $n_x$ denotes the number

---

[1] Only the reactions involving $H_2$, He, C, Si, and O as either a reactant or product are shown.

© 2015 RAS

**Table 2.** Mass fraction and fractional abundances of all species which are different from zero at the beginning of the relaxation phase. Numbers in parentheses are powers of 10.

| Element | mass fraction | fractional abundance |
|---------|---------------|----------------------|
| H       | 7.18(-1)      | 1                    |
| He      | 2.79(-1)      | 9.73(-2)             |
| $C^+$   | 7.13(-4)      | 8.27(-5)             |
| O       | 1.42(-3)      | 1.24(-4)             |
| Si      | 6.78(-5)      | 3.37(-6)             |

density of the considered atom/molecule and $n_{H,tot}$ the number density of all H-nuclei, i.e. $n_{H,tot} = n_H + n_{H^+} + n_{H^-} + 2 \times n_{H_2} + ...$ . For example, carbon is depleted by a factor of 4.3 typical for molecular cloud cores, resulting in a fractional abundance of $8.27 \times 10^{-5}$ compared to its cosmic abundance of $3.55 \times 10^{-4}$. For Si we have used a depletion factor of 0.9, i.e. only 10% of Si are left in the gas phase (Savage & Sembach 1996), whereas Flower et al. (2005) have assumed a depletion of 100%, i.e. no Si left in the gas phase. We note that this description of depletion does not take into account the freeze-out of e.g. CO onto dust grains and therefore might somewhat overpredict its abundance. However, we emphasise that the ratio of $R = \frac{n_{CO}}{n_{H_2}}$ of about 1.5 $\times 10^{-4}$ found in our simulations is in general in rather good agreement with observations (see Section 5.1 of the paper and Tielens 2013, for a recent review). A more detailed investigation of the effect of freeze-out – also in combination with cosmic rays – on CO is planned for a subsequent paper.

We point out that at the start of the (chemical) relaxation phase, which lasts for 500 kyr in our model, we assume that carbon exists in its ionised form, $C^+$. All other species not listed in Table 2 are initialized with a mass fraction of $10^{-40}$ in order to avoid division by zero. We emphasise that the mass fraction of electrons in every cell is calculated self-consistently from the abundances of all charged species listed in Table 1.

Finally, we note that the FLASH code in general uses mass fractions for the advection of different species, which requires a conversion into number densities, which are used by KROME. This is taken care of by the `-flash` option when setting up KROME (see below), which creates a suitable patch of code to be incorporated into FLASH.

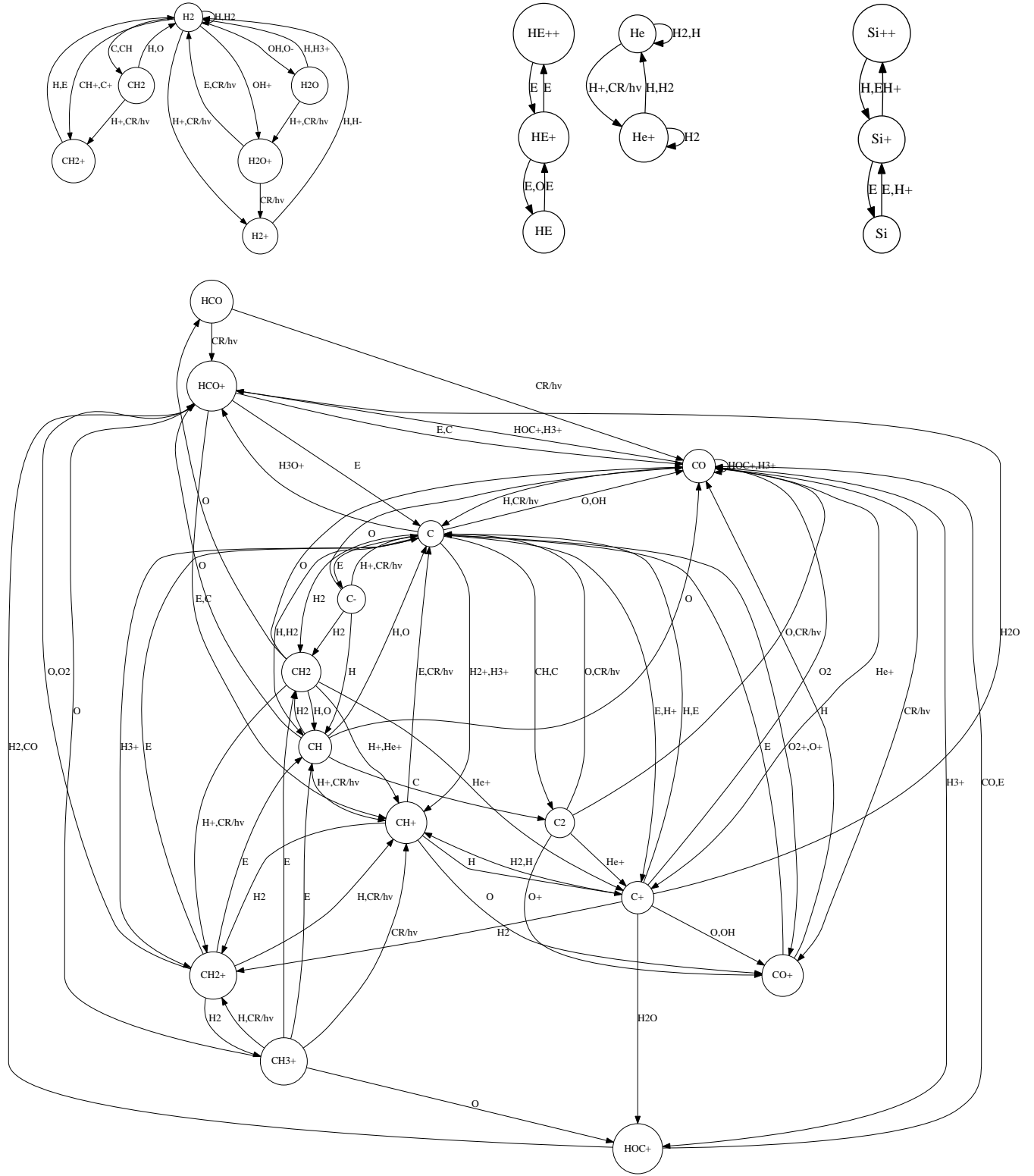

**Figure 1.** Reaction sub-sets involving different chemical species: $H_2$, (top left), helium (top middle), silicon (top right), carbon (bottom panel), and oxygen (next page). Each set only contains those reactions which include the chosen species as either a reactant or product. For this reason, the graphical representation is *not* complete, i.e. not all reactions contained in the network are included. The figure is created by the `pathway.py` script contained in the KROME package.

**Figure 1.** , continued. Reaction channels involving oxygen.

**Table 1.** List of the 37 chemical species considered in our network.

| | | | | | | | | | | | | | |
|---|---|---|---|---|---|---|---|---|---|---|---|---|---|
| $e^-$ | H | $H^+$ | $H^-$ | $H_2$ | $H_2^+$ | $H_3^+$ | He | $He^+$ | $He^{2+}$ | C | $C^+$ | $C^-$ | $C_2$ |
| CH | $CH^+$ | $CH_2$ | $CH_2^+$ | $CH_3^+$ | O | $O^+$ | $O^-$ | $O_2$ | $O_2^+$ | OH | $OH^+$ | $H_2O$ | $H_2O^+$ |
| $H_3O^+$ | HCO | $HCO^+$ | $HOC^+$ | CO | $CO^+$ | Si | $Si^+$ | $Si^{2+}$ | | | | | |

## 2 COOLING AND HEATING

Here we briefly describe all cooling and heating mechanisms used in our simulations. Most of the cooling and heating routines used come with the KROME package and can simply be incorporated by calling them during the setup of the code. The command with which the KROME code was prepared for usage within FLASH is given below (see also Grassi et al. 2014, for details of the compilations of the code):

```
./krome -n react_COthin -heating CHEM,CR,PHOTODUST
-cooling H2,CHEM,CIE,CI,CII,OI,OII,SiI,SiII,CO
-coolingQuench 10 -gamma FULL -useN -compact
-flash
```

We note that in the KROME package the gas temperature is updated simultaneously with the chemical abundances. The heating and cooling routines, which are directly adopted from the KROME package, are briefly described in the following. A more detailed description of these processes can be found in the KROME package (see also Grassi et al. 2014, for details). Only for those cooling routines which we modified or added by ourself, we give a more detailed description below.

## 2.1 Heating

The heating routines were linked in with the option `-heating CHEM,CR,PHOTODUST`. The option `CHEM` incorporates the heating due to formation of $H_2$ including the formation on dust grains. It also contains other exo- and endothermic processes related to the formation and destruction of $H_2$, which are linked with the `-cooling CHEM` flag. The option `CR` includes heating processes due to cosmic rays, which are caused by dissociation and ionisation of various species. This effect strongly depends on the strength of the chosen CRIR. The option `PHOTODUST` describes the heating of dust grains following Bakes & Tielens (1994) (see their equations 42 and 43). Here we emphasize that the strength as well as the attenuation of the ISRF has to be taken into account, which requires some slight modification of the code provided by KROME by setting the required strength of the ISRF to

$$G_0 \times \exp(-2.5 A_V), \quad (1)$$

where $A_V$ is the visual extinction, which is provided by the TreeCol algorithm (Clark et al. 2012; Wünsch et al. 2015, in prep.).

## 2.2 Cooling

The cooling routines used are compiled in by the usage of the options `-cooling H2,CHEM,CIE,CI,CII,OI,OII,SiI,SiII,CO` and `-coolingQuench 10`. The latter command causes the cooling to be reduced in case the gas temperature drops below 10 K.

The option `H2` describes cooling due collisional excitation of $H_2$ taken from Galli & Palla (1998) and Glover & Abel (2008). Further cooling effects due to $H_2$ dissociation are included with the option `CHEM`. Cooling due to collision-induced emission of $H_2$ is included via the `CIE` option following the approach of Ripamonti & Abel (2004). The cooling due to various metal lines is included via the options `CI,CII,OI,OII,SiI,SiII` and is described in detail in Maio et al. (2007).

The option `CO` includes the cooling due to CO line transitions. Here, some modifications have been made to the code provided by KROME. By default KROME uses a local approximation based on the density in the considered cell to calculate the CO column density, $N_{CO}$, which in turn is required to determine the CO cooling rate. We replace this approach by using $N_{CO}$, which is obtained self-consistently via the TreeCol algorithm (Clark et al. 2012; Wünsch et al. 2015, in prep.). In a second step we read out the actual CO cooling rate from a table provided by K. Omukai (based on the results of Neufeld & Kaufman (1993) and Omukai et al. (2010)), which contains the cooling rates depending on $N_{CO}$, the local hydrogen density, and the gas temperature.

Furthermore, in order to account for CO cooling in optically thick regions, we include cooling by the isotopes $^{13}CO$ and $C^{18}O$. In order to do so, in a first step we scale down $N_{CO}$ provided by the TreeCol algorithm by a factor of 69 and 557 for $^{13}CO$ and $C^{18}O$, respectively (Wilson 1999). We note that, as shown by Szűcs et al. (2014), for the density range considered in this work assuming the same abundance ratio for $^{12}CO$ and its isotopologues as for elemental carbon seems reasonable to first order. With the additional information of the local gas temperature and density we can then determine the cooling rate of both isotopes from the provided table which – since they depend on the $^{13}CO$ and $C^{18}O$ number density – have to be scaled down by the aforementioned factors.

Finally, we have to calculate the dust temperature in each cell. The dust temperature is required for an accurate determination of the $H_2$ formation rate on dust grains as well as the cooling of the gas via energy transfer to dust particles, which we also take into account as a gas cooling process (see below). The dust temperature is treated explicitly, i.e. it is only changed after each chemistry update, at which point it is set to the equilibrium temperature $T_{\rm dust,eq}$. This equilibrium temperature is determined by three different processes, the aforementioned energy transfer due to collisions between gas and dust particles, $\Lambda_{\rm coll}$[2], heating of the grains due to the incident ISRF, $\Gamma_{\rm ISRF}$, and cooling via black-body radiation, $\Lambda_{\rm BB}$:

$$\Lambda_{\rm coll}(T_{\rm dust,eq}) + \Gamma_{\rm ISRF}(T_{\rm dust,eq}) \stackrel{!}{=} \Lambda_{\rm BB}(T_{\rm dust,eq}) \quad (2)$$

First, $\Gamma_{\rm ISRF}$ is calculated by using the ISRF of Mathis et al. (1983) and the dust opacities from Ossenkopf & Henning (1994), which gives us

$$\Gamma_{\rm ISRF} = 5.8 \times 10^{-24} \chi\, n_{\rm H,tot}\, G_0 \,\rm erg\, s^{-1}\, cm^{-3}. \quad (3)$$

Here, $n_{\rm H,tot}$ is the number density of all hydrogen nuclei and $\chi$ the attenuation of the ISRF due to dust absorption, which depends on the column density of gas which in turn is provided by the TreeCol algorithm. For more details on the calculation of $\chi$ we refer the reader to the appendix in Glover & Clark (2012, see their equation A6).

The cooling of dust grains via black-body radiation is taken from Glover & Clark (2012) and given by

$$\Lambda_{\rm BB}(T_{\rm dust}) = 4.68 \times 10^{-31}\, T_{\rm dust}^6\, n_{\rm H,tot} \,\rm erg\, s^{-1}\, cm^{-3}. \quad (4)$$

For the heat transport from gas to dust via collisions we adopt the expression from Goldsmith (2001)

$$\Lambda_{\rm coll}(T_{\rm dust}) = 2 \times 10^{-33}\, n_{H_2}^2\, \sqrt{\frac{T_{\rm gas}}{10.0}}\, (T_{\rm gas} - T_{\rm dust}) \,\rm erg\, s^{-1}\, cm^{-3}. \quad (5)$$

Since this latter process also affects the gas temperature, we included it into the code provided by KROME via a user defined *gas* cooling routine. This routine, together with the cooling routines defined in the setup command given in the beginning of this section, is evaluated during the chemistry update and thus directly affects the gas temperature. Finally, we note that KROME is designed in a way that the gas temperature and the chemical abundances are updated simultaneously.

Finally, in Fig. 2 we show the dependence of $T_{\rm dust}$ on the gas density $\rho$ for the six simulations using the averaged data shown in Fig. 3 of the paper. In addition, we show a polytropic relation $T_{\rm dust} \propto \rho^{\gamma-1}$. As can be seen, the thermal behaviour of the dust is reasonable well described by the aforementioned polytropic relation with $\gamma$ between 0.9 and 0.95 (black and cyan line, respectively).

---

[2] We use the symbol $\Lambda$ since it is a *gas cooling* process.

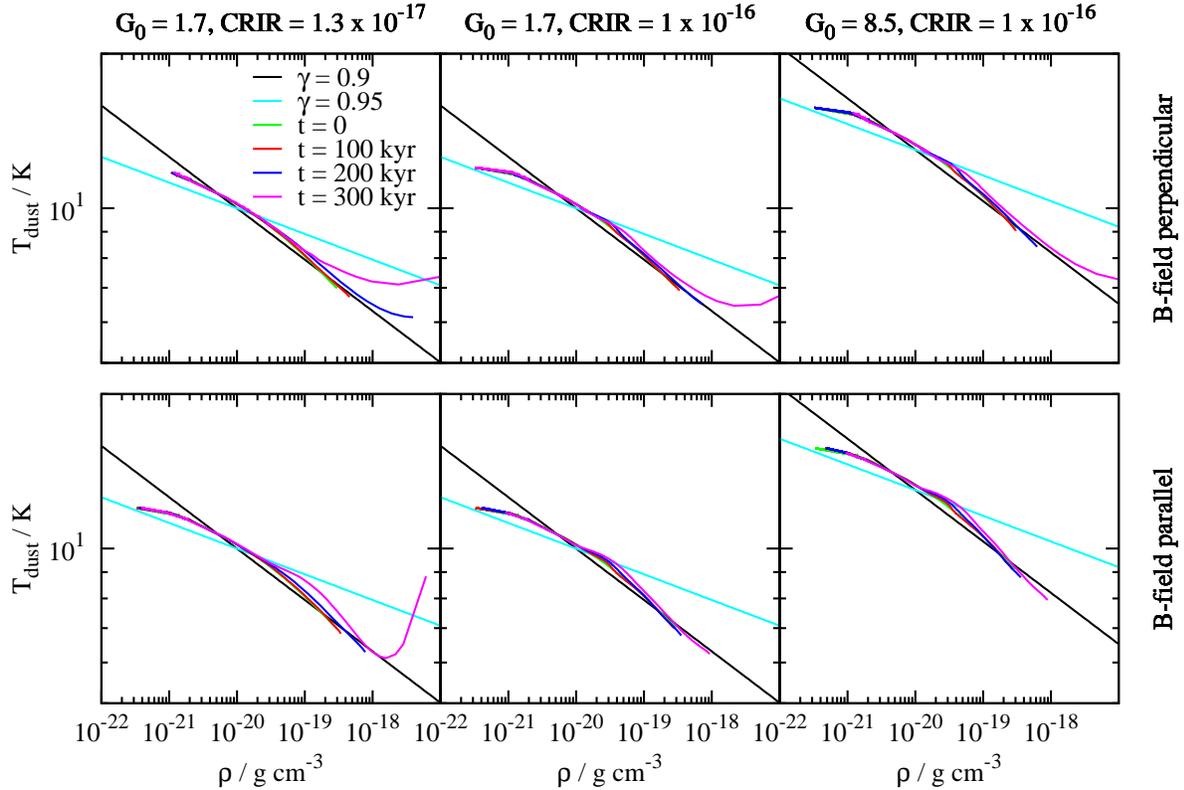

**Figure 2.** Dependence of the dust temperature on the gas density for all 6 six simulations considered in this work at t = 0, 100, 200, and 300 kyr. The top row shows the runs with a perpendicular magnetic field, the bottom row the runs with a parallel magnetic field. Note that at t = 0, the situation corresponds to that after an initial relaxation phase of 500 kyr. The straight lines show a polytropic relation $T_{\rm dust} \propto \rho^{\gamma-1}$ with $\gamma = 0.9$ and 0.95 (black and cyan line, respectively).

## 3 BENCHMARK TEST

In the following we compare our chemical model to a more sophisticated chemical model used by Röllig et al. (2007) in a benchmark test comparing different PDR codes. In detail we compare the results of our model to that produced with the KOSMA-$\tau$ code (Stoerzer et al. 1996), which were produced using a network which contains about 4000 reactions and 31 species. We use the tests denoted as F1 and F4 in Röllig et al. (2007), which are essentially 1-dimensional PDR models under the assumption of a fixed gas temperature of 50 K. The models consider a medium with a homogenous (total) hydrogen density of $10^3$ cm$^{-3}$ (F1) and $10^{5.5}$ cm$^{-3}$ (F4), which is irradiated by a UV field with a strength of 10 (F1) and $10^5$ times the Draine (1978) field. For both models self-shielding is taken into account and the CRIR is set to $5 \times 10^{-17}$ s$^{-1}$. For more details we would like to refer the reader to Röllig et al. (2007).

In Fig. 3 we show the results of the comparison for model F1 (left panel) and model F4 (right panel). In general there is a good agreement between our chemistry network and that used by Röllig et al. (2007). The transition from H to H$_2$ as well as that from C$^+$ to CO occur at similar AV values for both codes/networks. In general, also the abundances calculated with our reduced network compare reasonably well with that used in the KOSMA-$\tau$ code. Moreover, our results also clearly indicate that our shielding implementation within the TreeCol algorithm (Clark et al. 2012; Wünsch et al. 2015, in prep.), which is essential to reproduce the aforementioned transition from H to H$_2$ and from C$^+$ to CO, is working fine.

We emphasize that the differences seen between both codes are of similar magnitude than those differences appearing between the different codes presented in Röllig et al. (2007). Moreover, due to ongoing research over the last decade there are naturally some differences in the various rate coefficients present, which also lead to slightly different abundances. For this reason we argue that our reduced network gives a very good representation of significantly more extensive – for the work presented here actually too extensive – networks.

6

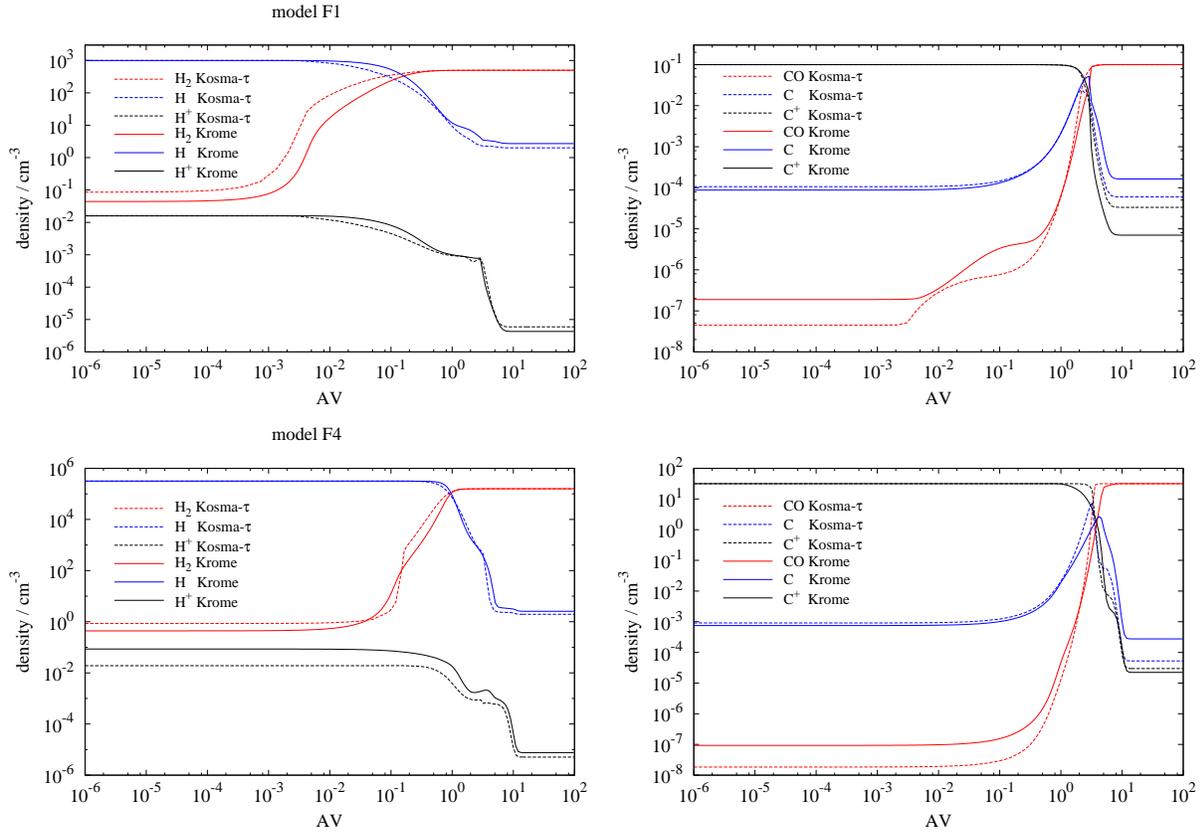

**Figure 3.** Results of two PDR benchmark tests F1 and F4 (see text) presented in Röllig et al. (2007) using the network presented in this work (Krome) and that of the KOSMA-$\tau$ code used in Röllig et al. (2007). In general we find a very good agreement between both codes despite the large differences in the network size.



\end{document}